\documentclass[amsmath, amssymb, aps, twocolumn]{revtex4-1}

\usepackage{graphicx}
\usepackage{booktabs}

\usepackage{txfonts}

\begin{document}

\title{Pressure Study of the Noncentrosymmetric 5$d$-Electron Superconductors Ca$M$Si$_3$~($M$= Ir, Pt)}

\author{Gaku EGUCHI$^1$}
\email{geguchi@scphys.kyoto-u.ac.jp} \author{Friedrich KNEIDINGER$^2$}  \author{Leonid SALAMAKHA$^2$} \author{Shingo YONEZAWA$^1$} \author{Yoshiteru MAENO$^1$} \author{Ernst BAUER$^2$}

\affiliation{$^1$Department of Physics, Graduate School of Science, Kyoto University, Kyoto 606-8502, Japan \\ $^2$Institute of Solid State Physics, Vienna University of Technology, A-1040 Wien, Austria}

\begin{abstract}
We report a hydrostatic pressure study on the Rashba-type noncentrosymmetric superconductors Ca$M$Si$_3$ ($M$= Ir, Pt). The temperature dependence of the resistivity of both compounds is well described by the conventional Bloch-Gr{\"u}neisen formalism at each pressure. This fact suggests that electron-phonon scattering is dominant in these compounds. The superconducting critical temperature $T_{\rm{c}}$ decreases with pressure above 0.41 to 2~GPa at a rate of $\sim0.2$ K/GPa for both compounds. This $T_{\rm{c}}$ behavior can be explained with a modest decrease in the density of states based on the conventional BCS theory.
\begin{description}
\item[Keywords]
noncentrosymmetric superconductor, hydrostatic pressure, $d$-electron system, resistivity, superconducting $H$-$T$ phase diagram
\end{description}
\end{abstract}

\maketitle

\section{Introduction}
Superconductors without the inversion symmetry have potentials to exhibit various novel phenomena and have recently been actively studied~\cite{Bauer2004PRL,Akazawa2004JPSJ,Kimura2005PRL,Nishiyama2007PRL,Shamsuzzaman2010JPCO,Harada2010PhysC,Peets2011PRB,Kase2009JPSJ044710,LeiFang2009PRB,Klimczuk2006PRB_Mg10Ir19B16,Tahara2009PRB_Mg10Ir19B16,Bauer2009PRB,Bauer2010PRB_Mo3Al2C,Karki2010PRB_Mo3Al2C,Fujimoto2007JPSJ051008,Oikawa2008JPSmeeting, Eguchi2011PRB}. Studies of this new category of superconductors, which are called noncentrosymmetric superconductors (NCSCs), were initiated by the discovery of superconductivity in the heavy Fermion CePt$_3$Si~\cite{Bauer2004PRL}. The upper critical field $H_{\rm{c2}}$ of this compound is substantially larger than the Pauli limiting field, at which the ordinary spin-singlet superconductivity becomes unstable owing to the Zeeman energy. This large $H_{\rm{c2}}$ is attributable to a mixing between spin-singlet and spin-triplet states resulting from the fact that the parity is no longer a meaningful label for NCSCs. In addition to such parity violation, another important feature of NCSCs is the existence of the anisotropic spin-orbit interaction (ASOI), which may affect both the normal and superconducting states. Up to now, a number of NCSCs have been discovered and investigated, e.g., UIr~\cite{Akazawa2004JPSJ}, CeRhSi$_3$~\cite{Kimura2005PRL}, CeIrSi$_3$~\cite{Sugitani2006JPSJ}, CeCoGe$_3$\cite{Settai2007IJMPB}, Li$_2$(Pd$_{1-x}$Pt$_{x}$)$_3$B~\cite{Nishiyama2007PRL,Shamsuzzaman2010JPCO,Harada2010PhysC,Peets2011PRB}, Ru$_7$B$_3$~\cite{Kase2009JPSJ044710,LeiFang2009PRB}, Mg$_{10}$Ir$_{19}$B$_{16}$~\cite{Klimczuk2006PRB_Mg10Ir19B16,Tahara2009PRB_Mg10Ir19B16}, BaPtSi$_3$~\cite{Bauer2009PRB}, Mo$_3$Al$_2$C~\cite{Bauer2010PRB_Mo3Al2C,Karki2010PRB_Mo3Al2C}, and Ca$M$Si$_3$ ($M$= Ir, Pt)\cite{Eguchi2011PRB}. Indeed, some of them exhibit unconventional behaviors: e.g., CeRhSi$_3$ and CeIrSi$_3$ have exceptionally high $H_{\rm{c2}}$ values, Li$_2$Pt$_3$B exhibits a temperature-independent Knight shift through the superconducting transition, and Mo$_3$Al$_2$C shows a non-BCS behavior in the specific heat and the nuclear-lattice relaxation rate 1/$T_1$. However, a majority of the NCSCs behave conventionally. Theoretically, it is predicted that the existence of a strong electron correlation in addition to a strong ASOI is crucial for unconventional superconducting phenomena~\cite{Fujimoto2007JPSJ051008}. In fact, this prediction seems to explain the reported unconventional behavior in the cerium-based NCSCs, in which strong electron correlations originate from the interaction of conduction electrons with $f$-electrons. However, the cerium-based NCSCs reported up-to-date also exhibit antiferromagnetic ordering within or near the superconducting phase. This fact complicates the situation, because antiferromagnetic spin fluctuation without parity violation can lead to non-$s$-wave Cooper pairing as in the cuprate superconductors. Thus, in order to extract the roles of parity violation and ASOI, studies of NCSCs without spin fluctuations are valuable.

Ca$M$Si$_3$ ($M$=Ir, Pt) compounds have recently been reported to be nonmagnetic, fully gapped superconductors without a strong electron correlation~\cite{Oikawa2008JPSmeeting, Eguchi2011PRB}. Thus, these compounds may serve as model materials for investigations of phenomena originating from their noncentrosymmetric crystal structure with a strong ASOI~\cite{Kaczkowski2011JAC}, because these compounds are not affected by any complications due to magnetism. The application of pressure is useful because effects of their electronic state variations can be investigated without introducing impurities by chemical substitution. For example, one may expect changes in superconducting properties due to the appearance/disappearance of an ASOI-split pair of Fermi surfaces. Here, we report the hydrostatic pressure dependence of the resistivity of Ca$M$Si$_3$($M$=Ir, Pt) as well as its magnetic field dependence, and discuss the variation in superconducting behavior.

\section{Experimental Procedure}
The arc-melted polycrystalline samples used in this study are from the same batch as the crystals used in ref.~17. The samples were shaped into blocks with the dimension of approximately $0.8$~$\times$~$0.8$~$\times$~$2$~mm$^3$. The samples were placed in a Teflon capsule with a hydrostatic pressure medium (Daphne 7373) and mounted in a commercial piston-cylinder pressure cell (R\&D Support Co., Ltd.). The pressure inside was determined from the Curie temperature of HoCo$_2$~\cite{Hauser1998PRB}, which was mounted inside the capsule together with the samples. Resistivity measurements with a conventional four-probe a.c. bridge technique under hydrostatic pressure were performed in a $^3$He cryostat (Cryogenic Ltd.) down to 0.35~K. The excitation current was 10~mA, and the frequency was 13.8~Hz.

\section{Results and Discussion}

\begin{figure}
\centering
\includegraphics[width=8.5cm, clip]{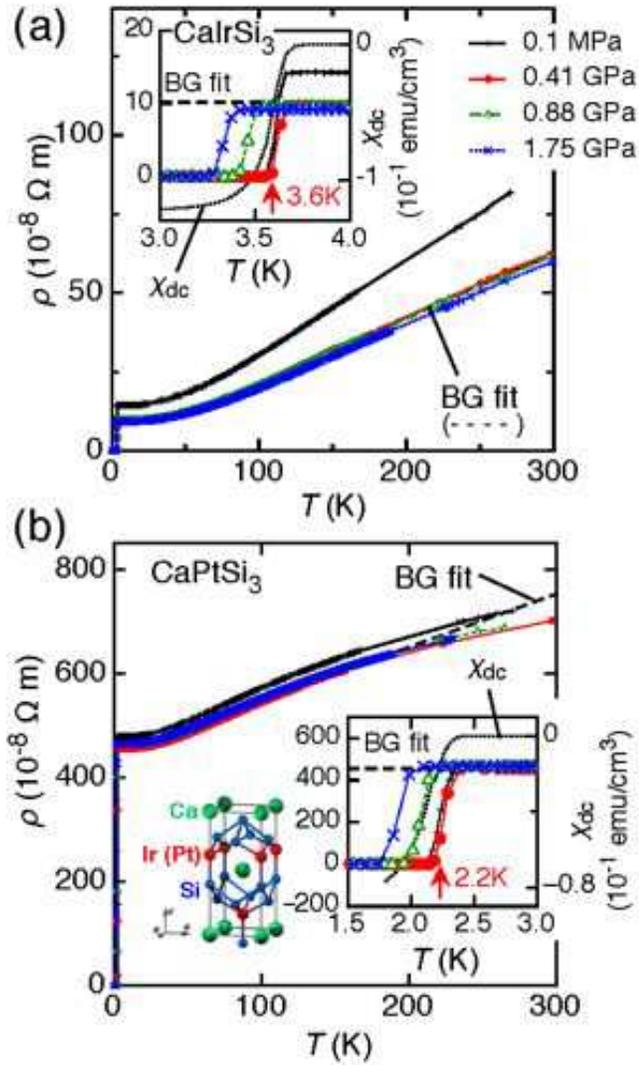}
\caption{(Color online) Temperature dependence of the resistivity of polycrystalline (a) CaIrSi$_3$ and (b) CaPtSi$_3$ for several pressures. Enlarged views near $T_{\rm{c}}$ are shown in each panel, and the zero-resistivity $T_{\rm{c}}$ values for 0~T and 0.41~GPa are indicated by arrows. The crystal structure is presented in the inset of (b). The fits of the resistivity for 0.41~GPa obtained by the Bloch-Gr{\"u}neisen (BG) formula are also presented. The insets also present the dc susceptibility of the samples from which the crystals used in the present pressure study were cut. The susceptibility was measured under zero-field-cooled condition at 1~mT under ambient pressure. These data indicate that the samples for the present pressure study exhibit an almost 100\% superconducting volume fraction.}
\label{rho_0T}
\end{figure}

The temperature dependence of resistivity at several pressures is presented in Fig.~\ref{rho_0T}, for which the sequence of the pressure application was in the order of 0.41~GPa, 0.88~GPa, 1.75~GPa, and 0.1~MPa. The superconducting transition temperature $T_{\rm{c}}$ is defined as the temperature at which the resistivity drops to 5\% of its normal-state residual resistivity $\rho_0$: $\rho/\rho_0 = 5\%$. $T_{\rm{c}}$ values below 0.41~GPa are almost identical to the value in vacuum reported in ref.~17, and decrease with pressure for both compounds.
Residual resistivity ratios $\rho_{300{\rm{K}}}/\rho_0$ under pressure are $\sim4$ for CaIrSi$_3$ and $\sim1.5$ for CaPtSi$_3$; these values are also consistent with the previous report~\cite{Eguchi2011PRB}. For both compounds, the resistivity at 0.1~MPa is larger than those at the other pressures. The difference does not seem to be intrinsic, i.e., it originats from the current path change by the pressure release. Each resistivity curve below 200~K is fitted by the conventional Bloch-Gr{\"u}neisen (BG) formula
\begin{align}
\rho=\rho_0+\Bigg(\frac{C}{\varTheta_{\rm{D}}^{}}\Bigg)\Bigg(\frac{T^5}{\varTheta_{\rm{D}}^5}\Bigg)\int_0^{\frac{\varTheta_{\rm{D}}}{T}}\frac{x^5}{(\rm{exp}[x]-1)(1-\rm{exp}[-x])} dx \notag
\end{align}
with three fitting parameters: $C$ is a temperature-independent material constant describing the electron-phonon interaction, $\varTheta_{\rm{D}}$ is the Debye temperature, and $\rho_0$ is the residual resistivity. 
The fitting for the 0.41~GPa data yields $\varTheta_{\rm{D}}=322$~K for CaIrSi$_3$, which is consistent with $\varTheta_{\rm{D}}=360$~K deduced from the specific heat below 5~K~\cite{Eguchi2011PRB}. This result suggests that the temperature dependence of resistivity is dominated by the electron-phonon scattering. However, we obtain $\varTheta_{\rm{D}}=171$~K for CaPtSi$_3$, which is substantially smaller than the value from the specific heat (370~K)~\cite{Eguchi2011PRB}. This difference in $\varTheta_{\rm{D}}$ indicates that phonons having lower frequencies mainly contribute to the electron-phonon scattering. In addition, the resistivity tends to saturate more strongly than the BG approximation in the high-temperature region, as often observed in inter-metallic compounds (e.g., ref.~22). 

\begin{figure}
\centering
\includegraphics[width=8.5cm, clip]{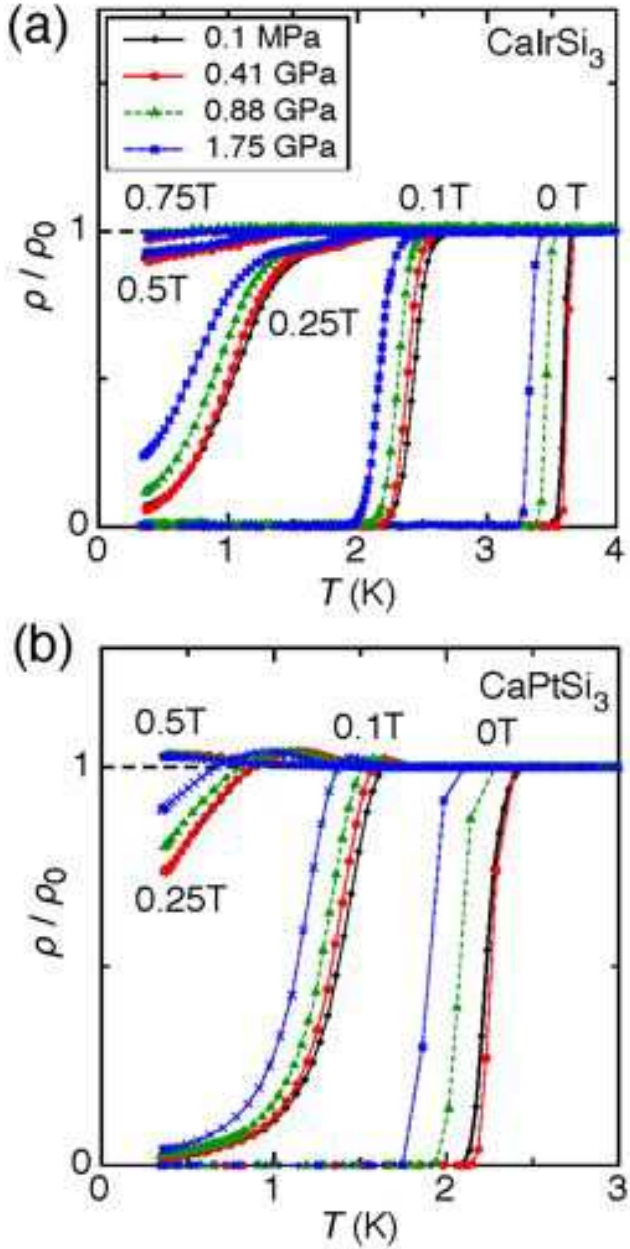}
\caption{(Color online) Magnetic field dependence of the low-temperature resistivity under pressure: (a) CaIrSi$_3$ and (b) CaPtSi$_3$. Each data is normalized by its residual resistivity $\rho_0$ obtained by Bloch-Gr{\"u}neisen fitting.}
\label{rho_fields}
\end{figure}

The magnetic field dependence of resistivity at several pressures is presented in Fig.~\ref{rho_fields}. The direction of the magnetic field was parallel to the applied current.  At each magnetic field, the overall temperature dependence for different pressures was similar to each other. This indicates that the electronic states as well as the scattering processes do not change much with pressure.
As presented in Fig.~\ref{rho_fields}, a clear field-dependent two-step transition was observed only for CaIrSi$_3$ which, additionally, exhibited a substantial broadening of the superconducting transition. This broadening indicates a spatial distribution of the superconducting fraction. Slight upturns just above the resistivity drop were observed for CaPtSi$_3$ in addition to the broadening of the transition. Such upturns have been observed and attributed to a pressure-induced charge-density-wave (CDW) ordering in Mo$_3$Sb$_7$~\cite{Tran2011JPCO}. However, in the case of CaPtSi$_3$, the upturn seems to be related to the superconducting transition because the onset temperature of this anomaly is suppressed by the field. Furthermore, the onset temperature is almost identical to the superconducting onset $T_{\rm{c}}$ reported in ref.~17. For these reasons, the upturns are unlikely associated with a CDW; rather, they are caused by the current path change or nontrivial vortex dynamics related to superconductivity.

\begin{figure}
\centering
\includegraphics[width=8.5cm, clip]{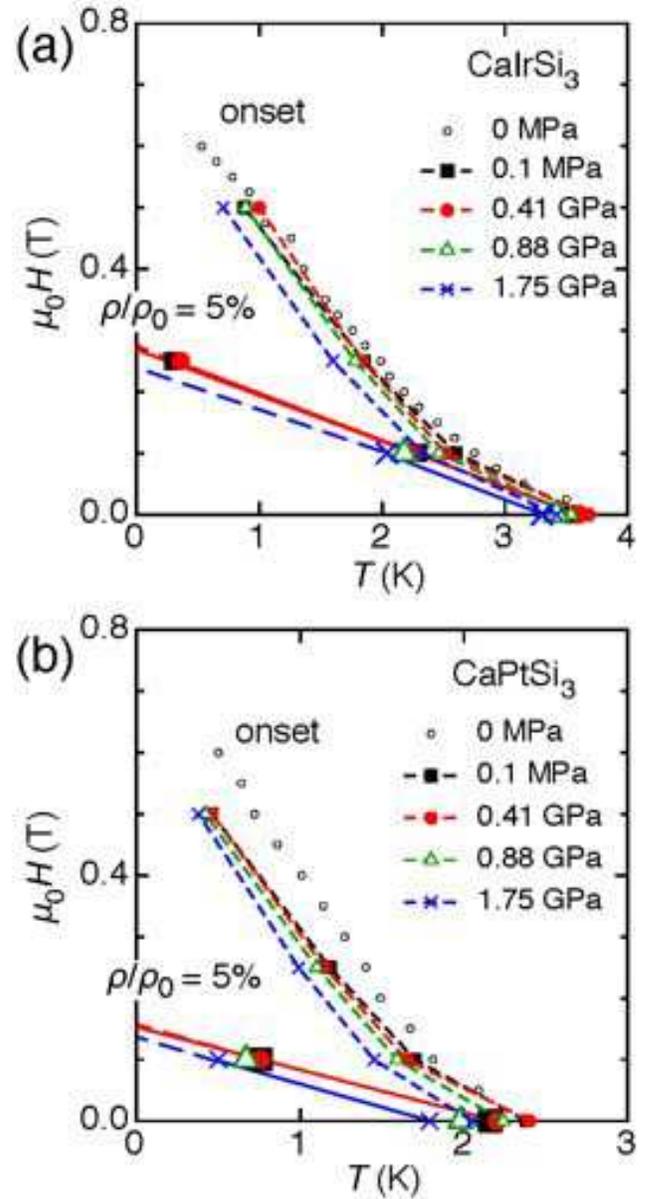}
\caption{(Color online) $H$-$T$ phase diagram under pressure: (a) CaIrSi$_3$ and (b) CaPtSi$_3$. The $H$-$T$ curves at 0~MPa are taken from the previous report~\cite{Eguchi2011PRB}. The large and small symbols indicate $T_{\rm{c}}$ and $T_{\rm{c}}^{{\rm{onset}}}$, respectively. See text for the definitions of $T_{\rm{c}}$ and $T_{\rm{c}}^{{\rm{onset}}}$.}
\label{HT}
\end{figure}

The $H$-$T$ phase diagrams deduced from the resistivity are presented in Fig.~\ref{HT}. Here, we show $T_{\rm{c}}$ defined as the 5\% criterion (large symbols) and $T_{\rm{c}}^{{\rm{onset}}}$ defined as the temperature at which the resistivity drops to 95\% of $\rho_0$ (small symbols). For CaPtSi$_3$ under magnetic fields, $T_{\rm{c}}^{{\rm{onset}}}$ is defined at the resistivity maximum in the transition region.
The large deviation in $T_{\rm{c}}$ from $T_{\rm{c}}^{{\rm{onset}}}$ reflects the broadening of the transition. $T_{\rm{c}}^{{\rm{onset}}}$ under zero pressure is taken from the previous report~\cite{Eguchi2011PRB}. The $H$-$T$ curves of the onset $T_{\rm{c}}$ under pressure have good overall similarity to those reported in ref.~17.

\begin{figure}
\centering
\includegraphics[width=8.5cm, clip]{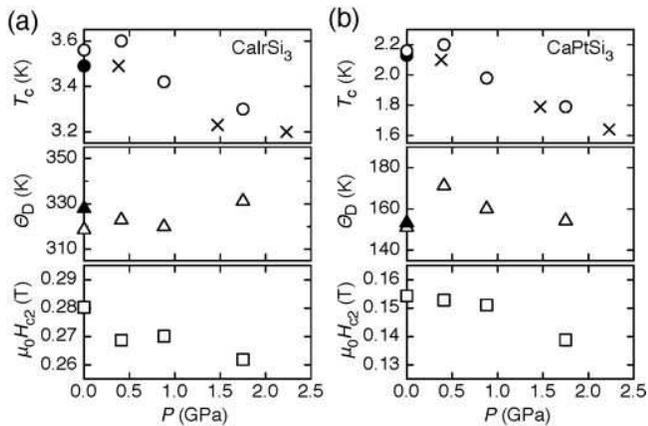}
\caption{Pressure dependence of superconducting transition temperature $T_{\rm{c}}$, Debye temperature $\varTheta_{\rm{D}}$, and the upper critical field $H_{\rm{c2}}(0)$ of (a) CaIrSi$_3$ and (b) CaPtSi$_3$, presented by open symbols. The values represented by the closed symbols are deduced from the data reported in ref.~17, and the $T_{\rm{c}}$ values represented by crosses are taken from a test measurement using the same samples.}
\label{rho_param}
\end{figure}

The pressure variations of $T_{\rm{c}}$, $\varTheta_{\rm{D}}$ determined by the Bloch-Gr{\"u}neisen formula, and $H_{\rm{c2}}(0)$ deduced from the linear extrapolation of $T_{\rm{c}}$ are presented in Fig.~\ref{rho_param}. A clear decrease in $T_{\rm{c}}$ is observed for both compounds above 0.41~GPa with a rate of approximately $dT_{\rm{c}}/dP \sim-0.2$~K/GPa. The variation in $T_{\rm{c}}$ below 0.41~GPa is within the experimental precision. In contrast, $\varTheta_{\rm{D}}$ does not exhibit any systematic change.

As discussed above, the electron-phonon and impurity scattering processes are dominant in both CaIrSi$_3$ and CaPtSi$_3$. No other interactions that could give rise to Cooper pairing, such as spin fluctuations, have been reported for these compounds. Therefore, pairing by the conventional phonon-mediated attractive interaction is most probably realized. Furthermore, specific heat is consistent with the weak-coupling BCS behavior~\cite{Eguchi2011PRB}. In the conventional weak-coupling BCS theory, $T_{\rm{c}}$ is given by $T_{\rm{c}}=1.13 \varTheta_{\rm{D}} \rm{exp}[-1/(N(0)V)]$, where $N(0)$ is the density of states at the Fermi level, and $V$ is the magnitude of the interelectron attractive interaction. Note that the formula indicates that $T_{\rm{c}}$ is an increasing function of $\varTheta_{\rm{D}}$, $N(0)V$. As mentioned above, the variation in $\varTheta_{\rm{D}}$ does not reflect the variation in $T_{\rm{c}}$ with pressure, and therefore should not dictate the observed pressure dependence of $T_{\rm{c}}$. 
As a result, the change in $T_{\rm{c}}$ with pressure likely originates from the change of $N(0)V$. In particular, the decrease in $T_{\rm{c}}$ with pressure above 0.41~GPa is probably due to the broadening of the bandwidth leading to the decrease in $N(0)$. 

$H_{\rm{c2}}(0)$ presented in Fig.~\ref{rho_param} tends to decrease with pressure.
This tendency is clearly observed from the raw data in Fig.~\ref{rho_fields}. The Ginzburg-Landau (GL) coherence length $\xi(0)$ values, calculated from the relation $\mu_0H_{\rm{c2}}(0)=\Phi_{\rm{0}}/2\pi\xi^2(0)$, where $\Phi_{\rm{0}}$ is the flux quantum, are  approximately 35~nm for CaIrSi$_3$ and 46~nm for CaPtSi$_3$ at 0.41~GPa with no significant change in the presented pressure range. This estimated $\xi(0)$ is an effective value given by $\xi=(1/\xi_0+1/l)^{-1}$, where $\xi_0$ is the intrinsic GL coherence length, and $l$ is the mean free path. Therefore, the change in effective coherence length can be explained by a change in either $\xi_0$ or $l$. Considering the fact that $l$ is determined by the mean interimpurity distance at low temperatures, $l$ should be independent of pressure. The pressure dependence of $\mu_0H_{\rm{c2}}$ is thus explained by the variation in $\xi_0$, which is reflected in the variation in $\xi$.

Our study indicates that $T_{\rm{c}}$ decreases by about 20\% with pressure (1.75~GPa) without changes in the behavior of the resistivity nor the shape of the $H_{\rm{c2}}(T)$ curve for both compounds. This indicates that the electronic states do not exhibit any substantial change in the examined pressure range, except for a modest decrease in $N(0)V$. However, the band calculations have revealed that several edges of ASOI-split bands are located in the vicinity of the Fermi energy~\cite{Kaczkowski2011JAC}. Thus, it is expected that one can induce a substantial change in the electronic structure if one can shift the band edges across the Fermi energy by applying higher pressure. Furthermore, when one Fermi surface of the ASOI-split pair disappears, a topological superconducting state might be realized. Indeed, the creation of a topological superconducting state by the disappearance of one part of ASOI-split Fermi surfaces has been proposed in a slightly different context~\cite{MSato2009PRB}. A study of the uniaxial pressure effect using a single crystal would also be favorable for investigating the relationship between the electronic state and the Rashba-type ASOI.

\section{Conclusions}
We have investigated the hydrostatic pressure dependence of the resistivity and the superconducting behavior of the noncentrosymmetric superconductors CaIrSi$_3$ and CaPtSi$_3$. The temperature dependence of resistivity is explained by the conventional Bloch-Gr{\"u}neisen formalism, suggesting that the electron-phonon scattering is predominant in these compounds. The decrease in $T_{\rm{c}}$ for $P > $~0.41~GPa up to the maximum pressure in the present study can be explained by the decrease in $N(0)V$ within the conventional BCS theory. The magnetic field dependence does not exhibit any qualitative change with pressure.

\section{Acknowledgment}
We thank D. C. Peets and M. Kriener for fruitful discussion. This work was supported by a Grant-in-Aid from the Global COE program ``The Next Generation of Physics, Spun from Universality and Emergence'' from the Ministry of Education, Culture, Sports, Science, and Technology (MEXT) of Japan, and by the ``Topological Quantum Phenomena'' Grant-in Aid for Scientific Research on Innovative Areas from MEXT of Japan. The work performed in Vienna was supported by the Austrian FWF, P22295. G.E. was supported by the Japan Society for the Promotion of Science (JSPS).

\end{document}